\documentclass[aps,english,twocolumn]{revtex4}
\usepackage{mathrsfs}
\usepackage{graphicx}
\usepackage{amsmath, amssymb}
\usepackage{babel}

\begin{document}

\title{Kondo Spin Screening Cloud in Two-dimensional Electron Gas with Spin-orbit Couplings}
\preprint{1}

\author{Xiao-Yong Feng$^{1}$ and Fu-Chun Zhang$^{2}$}
\affiliation{$^{1}$Department of physics, Normal University of Hangzhou, Hangzhou 310036, China\\$^{2}$Department of physics and Centre of Theoretical and Computational Physics,\\
 University of Hong Kong, Hong Kong, China}

\begin{abstract}
A spin-1/2 Anderson impurity in a semiconductor quantum well with
Rashba and Dresselhaus spin-orbit couplings is studied by using a
variational wave function method. The local magnetic moment is found
to be quenched at low temperatures. The spin-spin correlations of
the impurity  and the conduction electron density show anisotropy in
both spatial and spin spaces, which interpolates the Kondo spin
screenings of a conventional metal and of a surface of
three-dimensional topological insulators.
\end{abstract}
\maketitle

A spin-1/2 impurity with large on-site repulsion and occupation
energy well below the Fermi level of a conducting band will form a
local magnetic moment\cite{Anderson}. At low temperatures, the
impurity state hybridizes or couples with the conduction electron
states and the local magnetic moment is completely screened by a
conduction electron spin cloud surrounding the impurity, which is
known as the Kondo effect\cite{Kondo}.  To intuitively understand
the Kondo effect, we may consider N-1 conduction electron and one
impurity electron.  In the absence of the hybridization, the
impurity spin is free to form a local moment due to the on-site
Coulomb repulsion at the impurity site.  In the presence of a
small hybridization, the virtual process of the hopping between
the conduction and impurity states favors a spin singlet state of
the total N electron system.  More detailed Wilson's
renormalization group analysis shows that the energy gain is of
order of the Kondo temperature, which has an exponential
dependence to the coupling~\cite{book}.  After decades of
intensive study, the Kondo effect as a quantum impurity problem is
well understood\cite{book}. The Kondo spin screening cloud has
long been theoretically predicted. For more recent studies, see
for exmaple \cite{evidence1,evidence2,evidence3}. However, direct
experimental observation of the Kondo screening cloud turns out to
be more difficult. It was proposed that nuclear magnetic resonance
could be use to measure the spin polarization of conduction
electron spins of proper metallic sample in dilute
magnetic-impurities (Fe for example) by applying a small magnetic
field. The effect was later found too tiny to be detected
\cite{undetect}, due to the small energy scale or the large Kondo
spin cloud involved in the
screening\cite{proposal1,proposal2,proposal3,proposal4}. Recent
development on the spin-resolved scanning tunneling microscopy may
open a new possibility to detect the Kondo screening spin cloud in
experiments by directly measuring the correlation between the
impurity spin and the conduction electron spin density. This
revives our interest in study of Kondo spin screening cloud.

In a metal or a semi-conductor, the spin-orbit coupling breaks
spin SU(2) symmetry.  The spin of an electron feels an effective
momentum-dependent magnetic field and is no longer conserved. A
natural question is effect of the spin orbit coupling to the Kond
Problem. Recently, the Kondo screening of an Anderson impurity in
a helical metal has been studied\cite{we,Zitko}. The problem is of
interest in connection to the 3-dimensional (3D) topological
insulator, whose surface states are described by a helical metal
with Dirac dispersion. The magnetic doping on the surface of the
3D topological insulator $Bi_{2}Se_{3}$ has recently been
realized.\cite{magnetic doping}. Similar to the Kondo problem in a
conventional metal, the magnetic moment in a helical metal is
found to be completely quenched at low temperatures. However, the
texture of the spin correlations between the conduction electron
and the impurity become more complex. $\check{Z}$itko~\cite{Zitko}
was able to map the Anderson impurity model in a helical metal to
a conventional Kondo problem, in which the conduction electrons
carry pseudo-spins.  His work, however, does not provide direct
information about the spin screening, since the pseudo-spin is a
complicated combination of the real spins. The helical metal may
be considered as a strong spin-orbit coupling limit of a 2D
electron gas with Rashba spin-orbit coupling.  In a Rashba model,
the ratio of the spin-orbit coupling to the kinetic energy of the
conduction electron is a tunable parameter. Therefore, the model
allows us to study the Kondo spin screening continuously from a
conventional to a helical metal.

In this paper, we consider an Anderson impurity in a 2D electron
gas with Rashba and Dresselhaus spin-orbit couplings, which are
common in narrow-gap semiconductor quantum wells\cite{Rashba,
Dresselhaus}.  There have been a lot of activities recently on the
Rashba systems in connection with the semi-conductor spintronics.
The study of the Kondo screening in these systems may be of
interest in manipulation of the spins. We use a variational wave
function to study the Kondo screening. Similar to the Anderson
impurity models in conventional metal or in a helical metal, we
find that the magnetic moment is also completely quenched in a
semiconductor of Rashba coupling at low temperatures. The
spin-spin correlations between the impurity and the conduction
electron density are anisotropic in both spatial and spin spaces.
In addition to the diagonal components of the spin-spin
correlations, the off-diagonal spin-spin correlations have finite
values due to the broken SU(2) symmetry.

The system we study is described by a 2D Hamiltonian in $x-y$
plane. In Nambo's spinor representation, it reads
\begin{eqnarray}\label{H}
H &=& H_{c} + H_{h}+ H_{d}\\
H_{c}&=&\sum_{\mathbf{k}}c^{\dag}_{\mathbf{k}}
(\frac{\hbar^2\mathbf{k}^2}{2m^*}-\mu)c_{\mathbf{k}} +\nonumber
\\
&&c^{\dag}_{\mathbf{k}}[\alpha(\sigma_x k_y-\sigma_y
k_x)+\beta(\sigma_xk_x-\sigma_yk_y)]c_{\mathbf{k}} \nonumber\\
H_{h}&=&\sum_{\mathbf{k}}V_{\mathbf{k}}c^{\dag}_{\mathbf{k}}d+ h.c. \nonumber \\
H_{d}&=&(\epsilon_{d}-\mu)d^{\dag}d+Ud_{\uparrow}^{\dag}d_{\uparrow}d_{\downarrow}^{\dag}d_{\downarrow}.
\nonumber
\end{eqnarray}
In the above expressions, $\mathbf{k}=(k_x, k_y)$, and $H_c$ is
the Hamiltonian for the conduction electrons of effective mass
$m^*$. $\alpha$ and $\beta$ describe the strengths of the Rashba
and Dresselhaus type spin-orbit couplings, respectively. $\mu$ is
the chemical potential.
$c_{\mathbf{k}}^{\dag}=(c_{\mathbf{k}\uparrow}^{\dag},c_{\mathbf{k}\downarrow}^{\dag})$
is the creation operator in spinor representation. $\sigma_x$ and
$\sigma_y$ are the Pauli matrices. $H_d$ is the Hamiltonian for
the impurity states. The impurity is placed at the origin of the
x-y plane, and the energy level of the impurity state is
$\epsilon_d$, and
$d^{\dag}=(d_{\uparrow}^{\dag},d_{\downarrow}^{\dag})$. The
impurity state hybridizes with the conduction electrons through
$H_{h}$.

Let us first focus on the conduction electrons. The strength
of the Dresselhaus spin-orbit coupling is determined by the
asymmetric atomic field of the crystal lattice and it is absent in
the crystal with structural inversion symmetry. The strength of
the Rashba spin-orbit coupling in a quantum well structure is
allowed to be tuned by an external gate voltage and it is usually
one order in magnitude smaller than the kinetic energy of the electron gas.

After diagonalizing $H_c$, we obtain the single electron energy
eigenvalue $\epsilon_{\mathbf{k}s}$ and the corresponding
quasiparticle operators $\gamma_{\mathbf{k}}$,
\begin{eqnarray}\label{gamma}
\epsilon_{\mathbf{k}s} &=& \frac{\hbar^2k^2}{2m^*}-\mu+s\sqrt{(\alpha^2+\beta^2)k^2+4\alpha\beta k_xk_y} \nonumber \\
\gamma_{\mathbf{k}
s}&=&\frac{1}{\sqrt{2}}(e^{i\theta_{\mathbf{k}}/2}c_{\mathbf{k}\uparrow}+
ise^{-i\theta_{\mathbf{k}}/2}c_{\mathbf{k}\downarrow})
\end{eqnarray}
where $s=\pm$ is the index of the energy band,
$k=(k_x^2+k_y^2)^{\frac{1}{2}}$ and $\theta_{\mathbf{k}}$ is the
angle of the vector $(\alpha k_x+\beta k_y, \alpha k_y+\beta k_x)$
with the $k_x-$ axis. The ground state of many electron system
$H_c$ in the absence of $H_d$ is then simply given by,
$|\Psi_0\rangle=\prod_{\{\mathbf{k\pm}
\}\in\Omega}\gamma^{\dag}_{\mathbf{k}\pm}|0\rangle $, where
$|0\rangle$ is the vacuum, and $\Omega$ denotes the Fermi sea.

Now we consider the system with the impurity. For an isolated
impurity, if $\epsilon_d < \mu < \epsilon_d +U$ , the impurity is
singly occupied and a local magnetic moment is formed. Due to the
presence of $H_h$, the conduction electron hybridizes the impurity
state and tends to screen the impurity moment at low temperature.
To study the ground state properties of the system, we adopt the
trial wave-function method, which was first introduced by Varma
and Yafet in 1976\cite{wavefunction}.  The trial wavefucntion
captures many of the important aspects in Kondo problem, and has
been widely used in literature.  We believe the method should at
least qualitative or semi-quantitatively accurate to study the
Kondo spin screening cloud for our purpose. The trail wavefunction
is given\cite{we},
\begin{eqnarray}\label{wavefunction}
|\Psi\rangle=(a_{0}+\sum_{\kappa}
a_{\kappa}d^{\dag}_{\kappa}\gamma_{\kappa})|\Psi_0\rangle
\end{eqnarray}
where $\kappa=\{\mathbf{k},s\}$ and
$d_{\kappa}=\frac{1}{\sqrt{2}}(e^{i\theta_{\mathbf{k}}/2}d_{\uparrow}+
ise^{-i\theta_{\mathbf{k}}/2}d_{\downarrow})$. The large $U$ limit
is taken so that the double occupation of the impurity electrons
is completely excluded. The parameter $a_0$ and $a_\kappa$ are to
be determined using the variational method which optimizes the
ground state energy
$E=\langle\Psi|H|\Psi\rangle/\langle\Psi|\Psi\rangle$. Note that
the variational parameter $\kappa$ depends on both momentum and
spin. The spin-dependence is via the band index $s$ contained in
$\kappa$, which is a function of the momentum. The construction of
the trial wave-function (\ref{wavefunction}) is similar to that in
Ref.\cite{wavefunction} for a helical metal. In the limit
$\alpha=\beta=0$, the trial wavefunction is reduced to that for
Anderson impurity in conventional metal.  The wavefunction
contains the lowest energy sector of the single-particle
components and is expected to grasp the essential physics of the
Kondo problem. If all kinds of the particle-hole excitations of
the conduction electrons are included, the wave-function would  be
the exact eigenstate of the system.

The variational procedure leads to the expression for
$a_{\kappa}$,
\begin{eqnarray}\label{alpha}
a_{\kappa}=\frac{V_{\mathbf{k}}}{\epsilon_{\kappa}-\Delta_b}a_{0}
\end{eqnarray}
This will be used in the calculation of the spin correlations
between impurity and conduction electrons. The equation to
determine the binding energy, defined by the energy difference
between the state without the hybridization term $H_h$ and the
variational state, namely $\Delta_b =
\sum_{\kappa\in\Omega}\epsilon_{\kappa}+\epsilon_d-\mu-E$ is given
by,
\begin{eqnarray}\label{Delta}
\epsilon_d-\mu - \Delta_b =
\sum_{\kappa\in\Omega}\frac{V_{\mathbf{k}}^2}{\epsilon_{\kappa}-\Delta_b}.
\end{eqnarray}
$\Delta_b>0$
means that the variational state has lower energy and is stable
against the decoupled state.

In the case of $\mu\gg\frac{2m^*}{\hbar^2}(\alpha^2+\beta^2)$, we
can analytically solve Equation (\ref{Delta}) and find
\begin{eqnarray}
\Delta_b=\mu\exp\left[-\frac{\hbar^2}{2\pi
m^*V^2}(\mu-\epsilon_d)\right]
\end{eqnarray}
where we have assumed a momentum-independent hybridization
strength $V_{\mathbf{k}} = V$, namely we have assumed the
hybridization is on-site in the real space. a finite binding energy shows
the independent impurity plus conduction electron system is unstable against the
hybridization of the impurity and conduction electron with opposite spins described in Eq. (3).
Since the ground state is singlet, the local magnetic moment is quenched.

To investigate the role of the spin-orbit couplings to the Kondo
effect, we examine the magnetic properties of the system. The
central physical quantity we shall study is the correlation function
of the impurity spin
$\mathbf{S_d}(0)=\frac{1}{2}d^{\dagger}\boldsymbol{\sigma}d$ and the
spin density of the conduction electrons
$\mathbf{S_c}(\mathbf{r})=\frac{1}{2}c^{\dagger}(\mathbf{r})\boldsymbol{\sigma}c(\mathbf{r})$,
\begin{eqnarray}
J_{uv}(\mathbf{r})\equiv \frac{1}{\mathcal{N}}\langle\Psi|
S_{c}^{u}(\mathbf{r})S_{d}^{v}(0)|\Psi\rangle
\end{eqnarray}
where $\mathcal{N} = \sum_{\kappa\in\Omega}a_{\kappa}^2$ is the
possibility for there is a spin on the impurity and $u,v=x,y,z$ are
the spin indices. $J_{uv}$ have the same expression as in the Ref.
\cite{we}. We write them explicitly down here,
\begin{eqnarray}\label{JxJy}
J_{zz}(\mathbf{r}) &=& -\frac{1}{8}|\mathcal{B}(\mathbf{r})|^2+\frac{1}{8}|\mathcal{A}(\mathbf{r})|^2  \nonumber \\
J_{xx}(\mathbf{r})&=& -\frac{1}{8}|\mathcal{B}(\mathbf{r})|^2-\frac{1}{8}Re\mathcal{A}^2(\mathbf{r}) \nonumber \\
J_{yy}(\mathbf{r})&=& -\frac{1}{8}|\mathcal{B}(\mathbf{r})|^2+\frac{1}{8}Re\mathcal{A}^2(\mathbf{r}) \nonumber\\
J_{xy}(\mathbf{r}) &=& J_{yx} = -\frac{1}{8}Im\mathcal{A}^2(\mathbf{r})\nonumber\\
J_{xz}(\mathbf{r}) &=&
-J_{zx}=\frac{1}{4}Im(\mathcal{A}(\mathbf{r})\mathcal{B}(\mathbf{r}))
\nonumber\\
J_{yz}(\mathbf{r}) &=&
-J_{zy}=-\frac{1}{4}Re(\mathcal{A}(\mathbf{r})\mathcal{B}(\mathbf{r}))
\end{eqnarray}
In the above equations,
\begin{eqnarray}
\mathcal{A}(\mathbf{r})&=&\mathcal{N}^{-\frac{1}{2}}
\sum_{\kappa\in\Omega}se^{i(\mathbf{k}\cdot\mathbf{r}+\theta_{\mathbf{k}})}a_{\kappa},
\nonumber \\
\mathcal{B}(\mathbf{r})&=&\mathcal{N}^{-\frac{1}{2}}\sum_{\kappa\in\Omega}e^{i\mathbf{k}
\cdot\mathbf{r}}a_{\kappa}. \nonumber
\end{eqnarray}

\begin{figure*}
\includegraphics [width=14cm]{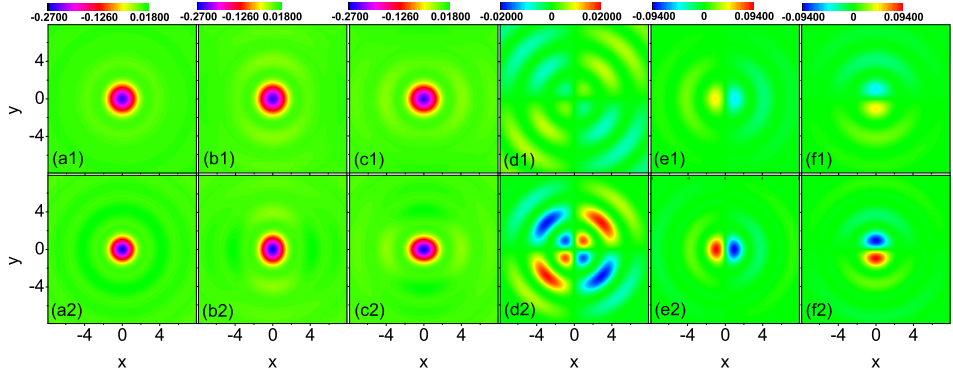}
\caption{(color online) Spatial distribution of the spin
correlations between impurity and conduction electron density for
systems with only Rashba spin-orbit coupling ($\beta=0$) (a1)-(f1)
are for $J_{zz}$, $J_{xx}$, $J_{yy}$, $J_{xy}$, $J_{xz}$ and
$J_{yz}$ with $\alpha = 0.3$; (a2)-(f2) are for the corresponding
correlations with $\alpha = 0.6$. The parameters are $\Delta_b=0.02$
and $\mu=1$. The length unit is $k_{0}^{-1}=10^{-9}$ meter, the
energy unit is $\frac{\hbar^2 k_{0}^2}{2m^*}=3.6\times10^{-2}eV$ and
$\alpha$ is in the unit of $\frac{\hbar^2 k_{0}}{2m^*}$.\cite{alpha}
}\label{correlations}
\end{figure*}

Firstly, we study the system with a pure Rashba coupling ($\alpha
\neq 0$, $\beta=0$). From the expression for $\mathcal{A}$ and
equation (\ref{alpha}), it is easy to check that
$\mathcal{A}(\mathbf{r})=0$ in the limit $\alpha=0$. In
Fig.\ref{correlations}, we plot the spatial distribution of all the
spin correlations between the impurity and the conduction electrons.
We see that the $J_{zz}$ is always isotropic about the origin point
while its density around this point decreases as $\alpha$ increases.
For $\alpha \neq 0$, $J_{xx}$ and $J_{yy}$ are anisotropic  and the
anisotropy becomes stronger at a larger $\alpha$, as we can see from
Fig. 2. In sharp contrast with the system without the spin-orbit
coupling, all of the off-diagonal spin correlations $J_{uv}$ where
$u\neq v$ are nonzero and they become larger for larger $\alpha$.
These results are closely related to the spin SU(2) symmetry broken.

Since the total spin of the conduction electrons is not a good
quantum number, the description of the spin singlet of the total
spin of the conduction electrons and the impurity spin is no
longer appropriate here. A proper quantity to measure the spin
part of the Kondo screening in the present systems is then the
correlation between the impurity spin and the total spin of the
conduction electrons, namely $I_0=\sum_{u}I_{u}$, where $u=x,y,z$
and $I_{u}=\int d\mathbf{r}J_{uu}(\mathbf{r})$.
$Re\mathcal{A}^2(\mathbf{r})$ has a $d$-wave symmetry in space and
it has no contribution to $I_x$ and $I_y$. We have $I_x=I_y$ and
usually they are not equal to $I_z$. Note that there is also the
orbital part of the Kondo screening, which contributes to the
screening of the impurity spin.  In the limit the Rashba coupling
(assumed to be positive without loss of generality) is small, we
have analytic expressions below for these correlations (hereafter
we use the units given in the caption of Fig. \ref{correlations}
for brevity),

\begin{eqnarray}\label{I}
I_z &\approx& -\frac{1}{4}(1-\frac{\alpha\sqrt{\mu}}{\Delta_b})\nonumber\\
I_x &=&
I_y\approx-\frac{1}{4}(1-\frac{\alpha\sqrt{\mu}}{2\Delta_b})\nonumber\\
I_0&\approx&-\frac{3}{4}(1-\frac{2\alpha\sqrt{\mu}}{3\Delta_b}).
\end{eqnarray}

\begin{figure}
\includegraphics [width=6cm]{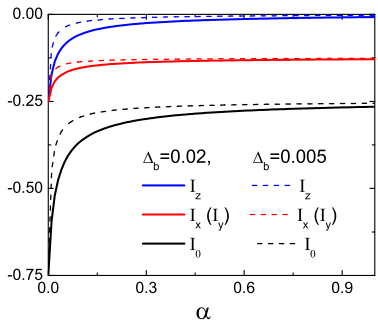}
\caption{(color online) Correlation functions, given in Eqn. (9),
between impurity spin and total spin of conduction electrons of the
ground state in Hamiltonian (1) with pure Rashba coupling. The solid
(dashed) lines are for $I_z$, $I_x(I_y)$ and $I_0$ with
$\Delta_b=0.02$ ( $\Delta_b=0.005$), and $\mu=1$. The energy and
length units are the same as in Fig. \ref{correlations}.}
\label{Rashba_screen}
\end{figure}
From the above equations, at $\alpha=0$, we have
$I_x=I_y=I_z=-\frac{1}{4}$ and $I_0=-\frac{3}{4}$, which is the
limiting case for the Kondo screening in conventional metal. For
more general values of the spin-orbit coupling, we show our
numerical results of  the correlations in Fig. 2.  As we can see,
$I_z$, $I_x$ or $I_y$ increases from $-1/4$, initially linearly at
small $\alpha$, then to a saturate value of $I_z=0$ or
$I_x=I_y=-1/8$ at large $\alpha$. Note that the value of $\alpha$
may be tunable in experiments by applying a gate voltage. We remark
that a small increase of $\alpha$ may suppress the spin correlation
significantly and this behavior is enhanced with smaller $\Delta_b$
as indicated in Eqn. (\ref{I}). Note that $I_z=0$ implies that
$z$-component of the impurity spin is unscreened globally by the
spins of the conduction electrons (As discussed in \cite{we}, the
$z$-component of the impurity spin is actually completely screened
by the $z$-component of the orbital angular momentum of the
conduction electron). The large $\alpha$ limit of $I_0$ is
$-\frac{1}{4}$ which is exactly the value for the Anderson impurity
in the helical metal with the chemical potential below the Dirac
point. Here we connect the Rashba system with the helical metal. The
helical metal exists on the surface of a 3-dimensional topological
insulator\cite{Fu1,Hsieh}. It is the large $\alpha$ (or large
$m^{*}$) limit of the Rashba system. For the helical metal, we have
to take an ultraviolet cutoff which is usually taken as the half
bulk gap. While for the Rashba system, to choose an energy cutoff is
not severe, the dispersion is bound below and the width of the
conduction band is a natural energy cutoff.

\begin{figure*}
\includegraphics [width=14cm ]{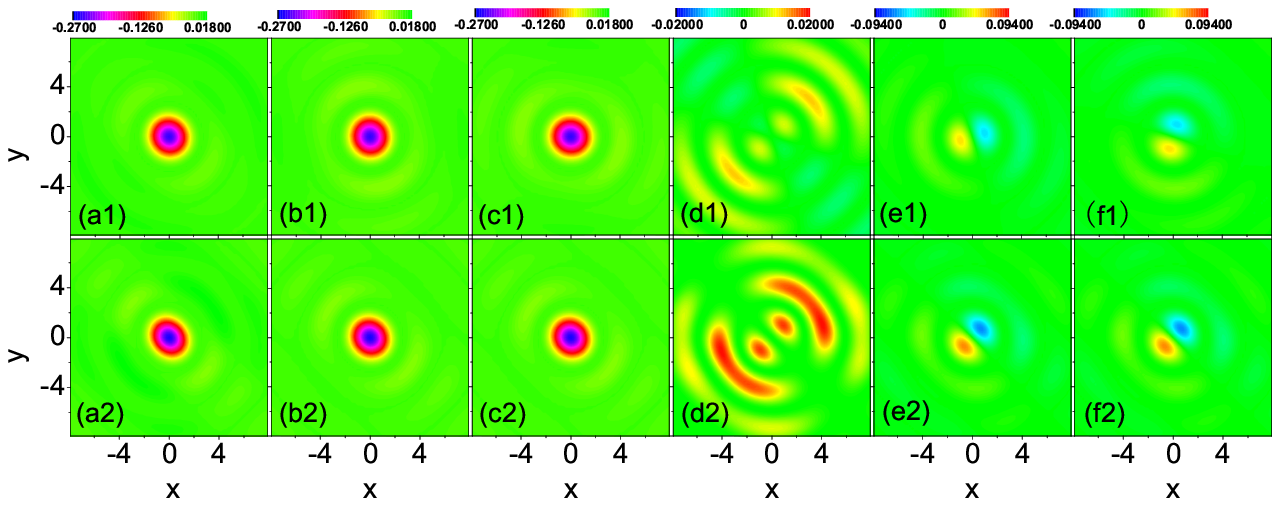}
\caption{(color online) Spatial distribution of the spin
correlations for systems with both Rashba and Dresselhaus spin-orbit
couplings. (a1)-(f1) are for $J_{zz}$, $J_{xx}$, $J_{yy}$, $J_{xy}$,
$J_{xz}$ and $J_{yz}$ with $\alpha = 0.3$ and $\beta = 0.1$;
(a2)-(f2) are for the corresponding ones with $ \alpha = \beta =
0.3$. The other parameters are $\Delta_b=0.02$ and $\mu=1$. The
energy and length units are the same as in Fig.
\ref{correlations}.}\label{RD}
\end{figure*}

Now we study the spin-spin correlation functions in the presence
of a pure Dresselhaus coupling. The problem can be mapped onto
pure Rashba coupling case by a unitary transformation as examined
by Shen et al.\cite{interchange}. Under the unitary transformation
$\mathcal{U}=\frac{\sqrt{2}}{2}(\sigma_x-\sigma_y)$, the Pauli
matrixes are transformed as $\sigma_x\rightarrow-\sigma_y$,
$\sigma_y\rightarrow-\sigma_x$ and $\sigma_z\rightarrow-\sigma_z$.
Therefore under such a transformation the Rashba and Dresselhaus
couplings in the Hamiltonian (\ref{H}) are
interchanged\cite{interchange}. If the system possesses pure
Dresselhaus coupling, it can be mapped onto the system with pure
Rashba coupling under the action of $\mathcal{U}$. Denote the spin
correlations of the Dresselhaus systems as $J_{uv}'$, then they
are related to the ones of the Rashba system as
$J_{zz}'(\mathbf{r})=J_{zz}(\mathbf{r})$,
$J_{xx}'(\mathbf{r})=J_{yy}(\mathbf{r})$,$J_{yy}'(\mathbf{r})=J_{xx}(\mathbf{r})$,$J_{xy}'(\mathbf{r})=J_{xy}(\mathbf{r})$,$J_{xz}'(\mathbf{r})=J_{yz}(\mathbf{r})$
and $J_{yz}'(\mathbf{r})=J_{xz}(\mathbf{r})$.

In the presence of both  the Rashba and Dresselhaus couplings, the
energy dispersion is no longer isotropic in momentum space as in the
Rashba system and so does the $a_{\kappa}$. However, they still
possesses reflection symmetries in momentum space about the lines
$k_x\pm k_y=0$. An immediate result is that $J_{zz}(\mathbf{r})$
loses its isotropy and is symmetric about $x\pm y=0$ in real space.
As shown in Fig. \ref{RD}, for systems with the Dresselhaus
coupling, the patterns of $J_{xx}$, $J_{yy}$, $J_{xz}$ and $J_{yz}$
are rotated and $J_{xy}$ is remarkably redistributed. At
$\alpha=\beta$, the model has an interchange symmetry between x and
y. In this case, $J_{xx}=J_{yy}$ and $J_{xz}=J_{yz}$.

In summary, we have examined a spin-1/2 Anderson impurity in a
two-dimensional electron system with spin-orbit couplings. The
binding energy and the spin correlation functions are calculated
using a trial wave-function method, which is widely used to study
Kondo problem. Being consistent with former
studies\cite{Yigal,Paaske}, the magnetic moment of the impurity is
found fully quenched at low temperatures, as the same as in the
conventional Anderson impurity problem. However, because of the
spin-orbit couplings, the spin correlations between the impurity and
the conduction electrons are different from the ones in conventional
metal. The diagonal components $J_{xx}$ and $J_{yy}$ are anisotropic
in space. The off-diagonal components are nonzero and have
particular spatial distributions. The introduction of the
Dresselhaus coupling changes the patterns of the correlation
functions dramatically.

Our theory is variational in nature. The variational wavefunction
method has been widely used to study Kondo impurity problem and
the basic results are consistent with more accurate methods such
as numerical renormalization group technique. We note that spatial
extension of the Kondo cloud in the  Anderson impurity model in
conventional metal was recently studied by Bergmann\cite{Bergmann}
by using Friedel artificially inserted resonance method, where the
screening length is quantitatively calculated. Our
focus in the present paper is on the anisotropic spin correlation
functions due to  the Rashba type spin-orbit
coupling. We expect that the spatial spin screening length
in our variational calculation is also associated with the Kondo
temperature, similar to Ref.\cite{Bergmann}.
In the parameter region we study, the screening decay is rather
rapid as we showed in the case of the Anderson impurity in a helical
metal\cite{we}.  We believe that our results presented here on the anisotropic
screening are qualitatively or semi-quantitative correct. Although more sophisticated methods may be needed to
provide precisely quantitative results of the Kondo spin cloud, it may be left for future work.
Our study may be of interest to semi-conductor spintronics, where the
Rashba system has played important roles. The impurity spin may be
used to control or to manipulate the conduction electrons and the
physics of Kondo screening in these systems should be interesting.
Direct experimental observation remains a major challenge. Nuclear magnetic resonance
 does not
seem to be promising for the tiny signal in large Kondo spin
cloud. Recently developed spin resolved STM may offer a new route.
By adding a small magnetic field around the impurity site, the STM
may in principle measure the spin-dependent local density of
states to probe both diagonal and off-diagonal spin correlations.
However, it may be still challenging to probe the electrons in the
quantum well structure, which may require long wave optical probe.
Nevertheless, the modern technique is developing very rapidly, and
the interesting phenomenon of the Kondo spin cloud, and the anisotropic spin screening examined in the present paper
will be observed in the near future.

We thank  Prof. S. Q. Shen for useful discussions. F. C. Zhang
acknowledges partial financial support from HKSAR RGC GRF grant
HKU701010.

\end{document}